\documentclass[sigconf]{acmart}
\usepackage{subcaption}
\usepackage{soul}
\usepackage{tipa}

\newenvironment{commentwrapper}[1]{\color{#1}}{\color{black}}

\definecolor{BLUE}{rgb}{0.0,0.2,0.7}
\definecolor{GREEN}{rgb}{0.0,0.7,0.2}
\definecolor{RED}{rgb}{0.8,0.0,0.0}

\newif\ifannotated

\ifannotated
    \newcommand{\del}[1]{\textcolor{blue}{\sout{#1}}}
    \newcommand{\rnr}[1]{\textcolor{red}{#1}}
\else
    \newcommand{\del}[1]{}
    \newcommand{\rnr}[1]{#1}
\fi

\AtBeginDocument{%
  \providecommand\BibTeX{{%
    \normalfont B\kern-0.5em{\scshape i\kern-0.25em b}\kern-0.8em\TeX}}}

\copyrightyear{2025}
\acmYear{2025}
\setcopyright{acmlicensed}
\acmConference[CHI '25]{CHI Conference on Human Factors in Computing Systems}{April 26-May 1, 2025}{Yokohama, Japan}
\acmBooktitle{CHI Conference on Human Factors in Computing Systems (CHI '25), April 26-May 1, 2025, Yokohama, Japan}
\acmDOI{10.1145/3706598.3714065}
\acmISBN{979-8-4007-1394-1/25/04}

\begin{document}

\title[MaRginalia]{MaRginalia: Enabling In-person Lecture Capturing and Note-taking Through Mixed Reality}

\author{Leping Qiu}
\orcid{0000-0003-0564-0456}
\affiliation{
  \institution{University of Toronto}
  \city{Toronto}
  \state{Ontario}
  \country{Canada}
}
\email{leping@dgp.toronto.edu}

\author{Erin Seongyoon Kim}
\orcid{0009-0007-5679-6895}
\affiliation{
  \institution{University of Toronto}
  \city{Toronto}
  \state{Ontario}
  \country{Canada}
}
\email{erinn.kim@mail.utoronto.ca}

\author{Sangho Suh}
\orcid{0000-0003-4617-5116}
\affiliation{
  \institution{University of Toronto}
  \city{Toronto}
  \state{Ontario}
  \country{Canada}
}
\email{sangho@dgp.toronto.edu}

\author{Ludwig Sidenmark}
\orcid{0000-0002-7965-0107}
\affiliation{
  \institution{University of Toronto}
  \city{Toronto}
  \state{Ontario}
  \country{Canada}
}
\email{lsidenmark@dgp.toronto.edu}

\author{Tovi Grossman}
\orcid{0000-0002-0494-5373}
\affiliation{
  \institution{University of Toronto}
  \city{Toronto}
  \state{Ontario}
  \country{Canada}
}
\email{tovi@dgp.toronto.edu}

\renewcommand{\shortauthors}{Qiu et al.}

\begin{abstract}
Students often take digital notes during live lectures, but current methods can be slow when capturing information from lecture slides or the instructor’s speech, and require them to focus on their devices, leading to distractions and missing important details. This paper explores supporting live lecture note-taking with mixed reality (MR) to quickly capture lecture information and take notes while staying engaged with the lecture. A survey and interviews with university students revealed common note-taking behaviors and challenges to inform the design. We present MaRginalia to provide digital note-taking with a stylus tablet and MR headset. Students can take notes with an MR representation of the tablet, lecture slides, and audio transcript without looking down at their device. When preferred, students can also perform detailed interactions by looking at the physical tablet. We demonstrate the feasibility and usefulness of MaRginalia and MR-based note-taking in a user study with 12 students.

\end{abstract}



\begin{teaserfigure}
  \includegraphics[width=\textwidth]{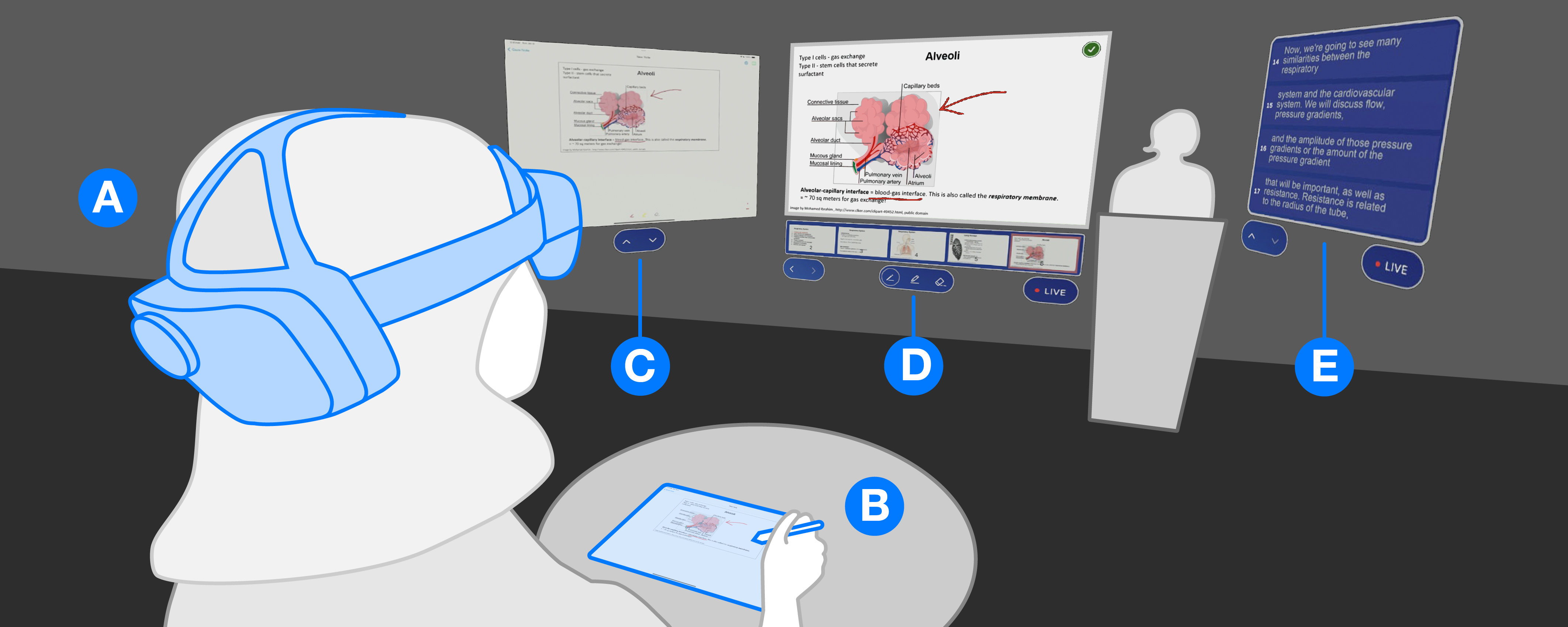}
  \caption{We present~MaRginalia, a novel mixed reality (A) and tablet (B) based note-taking system that enables users to take handwritten notes on the tablet (B) and in the MR spatial panel (C). The system automatically creates snapshots of the lecture slides and the lecturer's words to display on spatial panels for users to review (D and E). The user can quickly add the snapshots to their notes or annotate directly on the slide and transcripts with their annotation automatically added to their notes.}
  \label{fig:teaser}
\end{teaserfigure}

\maketitle

\section{Introduction}

\rnr{University students} take notes during lectures to encode new information and create external storage for review~\cite{divestaListeningNoteTaking1972}. With the increasing availability of tablets and laptops, more students take digital notes~\cite{grahameDigitalNoteTakingDiscussion2016}. Compared with traditional paper note-taking, digital note-taking \rnr{with tablets and laptops} offers more features, such as annotating slide decks, attaching photos taken in lectures, and integrating audio recordings. However, digital note-taking can introduce distractions from multitasking~\cite{kayExploringBenefitsChallenges2011}. Students may focus on non-learning activities, such as capturing and manipulating images, diverting attention from the lecture and leading to missing important information. \rnr{In addition, context switching between different sources of information might be detrimental to learning due to the split-attention effect~\cite{swellerSplitAttentionEffect2011}.}

Mixed reality head-mounted displays (MR HMDs) offer \rnr{the ability to seamlessly blend digital information into the physical world, which reduces distractions from context switching and enables} users to stay engaged with the task. Previous work explored augmenting physical objects, such as documents, to aid productivity by providing additional information to maintain user situational awareness and reduce context switching between different devices~\cite{liHoloDocEnablingMixed2019}, or summarizing and extracting relevant document information~\cite{gunturuRealitySummaryOnDemandMixed2024}. The front-facing camera and microphone can also capture the environment from the user's perspective. Prior work explored supplementing the user's document annotation with video clips from the HMD~\cite{digioiaInvestigatingUseAR2022}. However, using MR to aid lecture note-taking, where information has to be retrieved from multiple temporal sources, such as slides or lecturer speech, has remained unexplored.

\rnr{This project focuses on undergraduate and graduate university students due to their extensive experience with note-taking in lectures, their familiarity with diverse digital tools, such as tablets and laptops, and the flexibility they often have in adopting digital lecture note-taking methods compared with younger students.} To identify the required features to support digital note-taking, we conducted a formative study surveying 45 university students and interviewing 12 survey respondents to explore their note-taking behavior and challenges. The difficulties identified include combining materials such as pictures and recordings from different devices and missing lecture content during note-taking. The results informed the system's design goals: 1) support existing digital note-taking practices, 2) minimize friction in recording and organizing lecture content, and 3) enhance engagement during the lecture. 

Guided by the design goals, we developed MaRginalia\footnote{\rnr{MaRginalia, pronounced as \textsecstress mA:rdZI\textprimstress neIli\textipa{@}}, means marks made in document margins, such as scribbles, comments, and annotations.},
an MR- and tablet-based note-taking system (\autoref{fig:teaser}\rnr{)}. Through the MR HMD, MaRginalia displays a virtual projection of the user's tablet adjacent to the lecture slides and a \textit{spatial panel} of the transcript adjacent to the speaker. Users can capture or directly annotate lecture slides and transcripts and take notes on the tablet projection. This enables users to quickly capture content and take notes without looking down or switching between devices, helping them stay focused on the lecture. In addition, users can take notes directly on the tablet, just as they usually would. Finally, MaRginalia allows users to navigate the lecture slides and transcript history to ensure that important information is not missed during the lecture and allows users to track back to previous content freely.
 
We explored the usability and utility of~MaRginalia with 12 students in a simulated lecture environment. Participants found the system features helpful and easy to use. In summary, we contribute 1) MaRginalia, a novel MR- and tablet-based note-taking system; 2) formative study results on university students' digital note-taking behaviors and challenges; and 3) insights from an exploratory study to inform the future design of MR-based digital note-taking.

\section{Related Work}
Our work builds on previous research on digital note-taking, MR productivity, and \rnr{gaze, touch, and pen} interactions.

\subsection{Note-taking and HCI}

Note-taking is an essential activity for learning, as it creates external storage of information for future access, helps students encode the provided information~\cite{kiewraNotetakingFunctionsTechniques1991}, and helps students stay focused~\cite{kaneWhomMindWanders2007}. As notes are commonly taken while processing new information, note-taking is a complex cognitive activity, requiring students to perform multiple steps to understand, select relevant information, and process them into notes~\cite{piolatCognitiveEffortNote2005}. Also, as note-taking is often performed during time-sensitive events, such as lectures or meetings, there is an urgency to record information before it is lost. As such, being able to take notes quickly is a high priority~\cite{piolatCognitiveEffortNote2005, ardenDigitalNotetakingLectures2024}. Digital note-taking has become more prevalent due to the increasing availability of tablets and laptops and allows digital storage of notes for easy access~\cite{ardenDigitalNotetakingLectures2024}. Previous research has studied various ways of taking digital notes. Longhand (handwritten) notes on a tablet have been shown to be more effective in learning as they encourage students to rephrase information in their own words rather than copy it verbatim through typing~\cite{bochNoteTakingLearning2005}, but may be slower and thus less ideal during fast-paced lectures~\cite{moreheadNotetakingHabits21st2019}. As such, to quickly capture information, students may opt to take pictures of slides during lectures to capture all the information while also helping to learn and remember information through visual memory~\cite{dittaWhatHappensMemory2023, wongTakeNotesNot2023}. However, capturing information through different channels can create additional workload for students as the information is spread across various devices~\cite{swearnginScrapsEnablingMobile2021}. In our work, we focus on tablet-based longhand note-taking combined with MR for information capture.

The HCI community has put significant effort into tools to help students capture, contextualize, and organize notes. These systems are designed with different approaches to aid learning and make note-taking more efficient, such as digitizing physical notes~\cite{wilcoxDynomiteDynamicallyOrganized1997, brandlNiCEBookSupportingNatural2010}, environmental information such as audio, and blackboard writing~\cite{kalnikaiteMarkupYouTalk2012, shinVisualTranscriptsLecture2015, xuSemanticNavigationPowerPointBased2023,caoVideoStickerToolActive2022, banerjeeSegmentingMeetingsAgenda2007, chenMeetScriptDesigningTranscriptbased2023}, organizing information from multiple data sources into a single document~\cite{swearnginScrapsEnablingMobile2021, brandlNiCEBookSupportingNatural2010, fouseChronoVizSystemSupporting2011}, adding additional context and information to notes~\cite{hinckleyInkSeineSituSearch2007, olsenScreenCrayonsAnnotatingAnything2004}, or supporting collaborative note-taking~\cite{kamLivenotesSystemCooperative2005} \rnr{and virtual reality note-taking~\cite{chenIVRNoteDesignCreation2019}}. To help users avoid missing time-sensitive information such as speech due to distractions or not being able to keep up, several works have investigated playback support of lecture videos~\cite{mavaliTimeTurnerBichronousLearning2024}, interactive visualizations to effectively navigate videos~\cite{zhaoNovelSystemVisual2017}, navigable transcriptions of speech history~\cite{sonItOkayBe2023}, or embedding pictures of lecture slides to notes~\cite{chenIVRNoteDesignCreation2019}. In MaRginalia, we build on these works together with MR to enable the capture and organization of lecture content, live playback during lectures, and annotation without having to look down so that users can focus on the lecture.

\subsection{Mixed Reality Productivity Tools}

MR enables the integration of digital information into the physical environment and has therefore been explored to increase productivity. For example, \emph{HoloDoc} augments physical documents with digital content to provide context, search results, and multimedia playback in digital space~\cite{liHoloDocEnablingMixed2019}, while \emph{RealitySummary} provides text extraction and summary displayed around the physical document~\cite{gunturuRealitySummaryOnDemandMixed2024}. Meanwhile, \emph{GazePointAR} combines gaze, HMD cameras, and voice to provide contextual information about the physical environment~\cite{leeGazePointARContextAwareMultimodal2024}. MR has also been shown to be an effective platform for capturing physical information that can be integrated with notes~\cite{digioiaInvestigatingUseAR2022, speicher360AnywhereMobileAdhoc2018, choRealityReplayDetectingReplaying2023}, or integrating digital notes in physical environments through smartphones~\cite{qianDuallyNotedLayoutAware2022}, tablets~\cite{suzukiRealitySketchEmbeddingResponsive2020}, or HMDs~\cite{krugCleARSightExploring2022}. Finally, researchers have investigated the integration of other devices, such as smartphones~\cite{zhuBISHAREExploringBidirectional2020, zhuPhoneInVREvaluationSpatial2024} and tablets~\cite{suraleTabletInVRExploringDesign2019}, with HMDs to enable immersive yet precise interaction. These works highlight the benefits of MR in its ability to capture physical information and augment productivity through spatial interactions. In MaRginalia, we build on these projects in the context of in-person lecture note-taking, using MR to support easy capture and note-taking.

\subsection{\rnr{Gaze, Touch, and Pen Interactions}}

\rnr{MR note-taking requires indirect interactions that enable users to manipulate the virtual \textit{spatial elements}, such as lecture slides, placed at a distance from their seated position.} Gaze in combination with touch or pen input on a surface has been proposed as an effective technique for spatial interactions, as it allows direct and indirect input with little range of motion~\cite{pfeufferGazetouchCombiningGaze2014, turnerEyePullEye2013}. Because people's eyes are used to guide their hands, using gaze in combination with hand input leverages the natural eye and hand coordination, where gaze can be used to effortlessly cover large distances and hand input is used for precise and expressive interaction~\rnr{\cite{pfeufferGazetouchCombiningGaze2014, bienerBreakingScreenInteraction2020}}. Prior works have investigated gaze and pen interaction indirect interaction on touch surfaces~\cite{pfeufferGazetouchCombiningGaze2014, pfeufferGazeShiftingDirectIndirectInput2015}, content transfer between devices~\cite{turnerEyePullEye2013} and indirect interaction in extended reality~\cite{bienerPoVRPointAuthoringPresentations2022, gessleinPenbasedInteractionSpreadsheets2020}. In MaRginalia, we leverage gaze to quickly move interaction between different spatial elements and pen input to enable accurate and efficient note-taking directly on the tablet or indirectly on the spatial elements.
\section{University Student Note-taking Behavior}

We conducted a formative study with university students, \rnr{including undergraduate and graduate students}, to identify opportunities \rnr{and design goals for a digital lecture note-taking system}. We first completed a survey on \rnr{university students' lecture} digital note-taking habits and challenges with \rnr{their existing methods}. We then conducted a follow-up semi-formal interview with a subset of the students focused on specific usages of features and challenges.

\subsection{Survey}

The survey aimed to understand students' \rnr{digital note-taking behavior during lectures}. We distributed the survey to students of a research university in North America. The survey included questions on the frequency of digital note-taking, the frequency of using different tools for digital note-taking, subject ratings of the usefulness of digital notes, and a free-form question for them to comment on the most painful aspects of digital note-taking. Two survey respondents were randomly selected through a raffle to each receive a \rnr{Canadian Dollars CA\$50} gift card.

In total, 45 students (S1-S45) completed the survey \rnr{(21 women, 22 men, 2 non-binary)}, including 14 first- to third-year undergraduate students, 22 fourth-year or higher undergraduate students, and 9 graduate students. 77.8\% of them major in Engineering and Technology (35), 11.1\% in Math and Statistics (5), and the rest in Natural Sciences (3), Social Sciences (1), and Health Sciences (1).

\begin{figure*}
    \centering
    \includegraphics[width=1\linewidth]{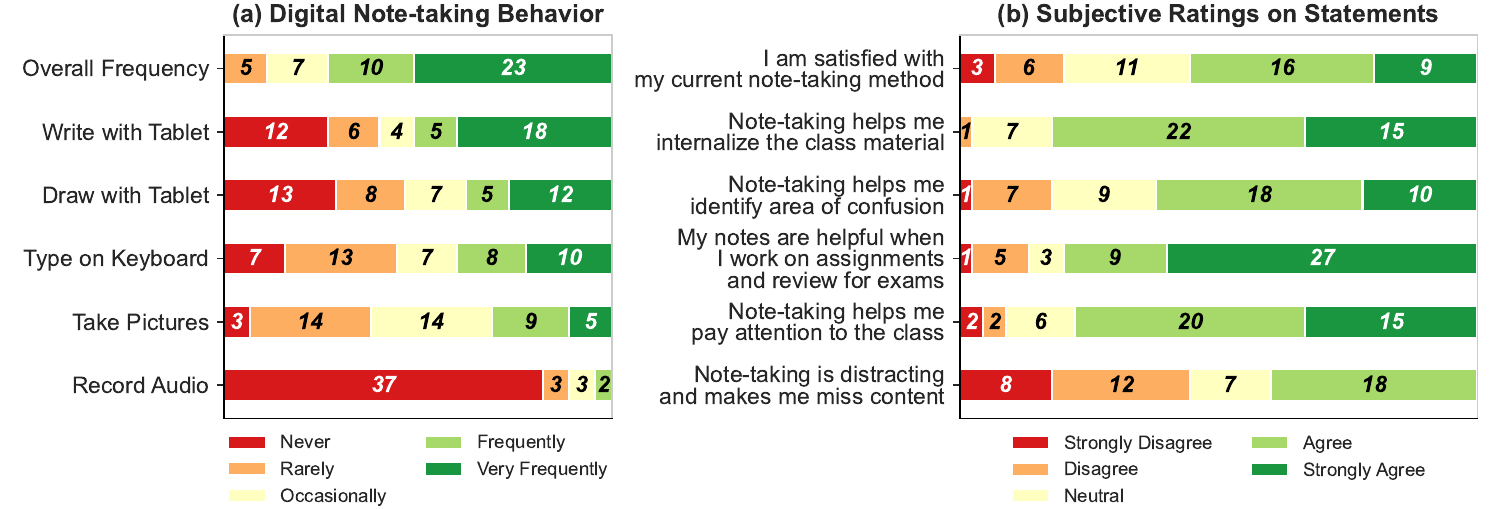}
    \caption{Survey results on digital note-taking \rnr{with 45 university students}. (a) The frequency of overall digital note-taking and specific \rnr{digital note-taking methods}. (b) Subjective ratings on \rnr{statements on respondents' digital note-taking satisfaction, utility, and issues based on their existing digital note-taking methods}.}
    \label{fig:survey-frequency}
\end{figure*}

\rnr{All survey respondents} take digital notes to varying degrees (\autoref{fig:survey-frequency}). Taking pictures is the most common \rnr{approach (42/45)}. Of the 11 students reported taking pictures frequently or very frequently, they also reported drawing and writing frequently or very frequently with tablets (10 and 9, respectively). The 13 students who do not use tablets mostly type notes, except \rnr{one student who reported solely taking pictures as digital note-taking}. Similar to the findings of a previous survey~\cite{ardenDigitalNotetakingLectures2024}, very few students reported using audio recordings (5/45). Of these, four reported frequently writing on tablets, and three reported frequently drawing on tablets. We did not find that the year of study and major affected the note-taking behavior. Overall, students are satisfied with their note-taking methods and find their notes especially useful in helping them internalize material, review, and pay attention to lectures. However, nine students are unsatisfied with their \rnr{current digital note-taking methods. Two frequently write and draw on tablets, five frequently type notes, and five frequently take pictures}. A significant proportion of survey respondents \rnr{(18/45)} consider taking notes to be distracting and make them miss content.

From the free-form responses, students' pain points with digital note-taking ranged from experiencing technical difficulties in the note-taking app and charging and carrying the devices (14/45), formatting notes such as typing and writing formulas (6/45), to eye fatigue (2/45). Two responded they had no pain points. More relevant to our project is 11 students shared that they missed content during lectures due to fast pacing. For example, S38 commented they ``\textit{can't catch up (with) the lecturer,}'' and S22 commented they missed content when they ``\textit{try to write more down while they (instructor) are speaking, distracting me from what they are saying.}'' Eight respondents shared that combining different sources of digital notes is cumbersome. For example, S26 commented: ``\textit{No good ways to combine different forms of content together (i.e., slide with annotations + text).}'' S30 commented: ``\textit{Syncing files between multiple devices}'' is the most painful aspect of digital note-taking, and S39 commented: ``\textit{It feels like it all spread out here and there: laptop, phone.}'' In addition, two students commented that their notes lacked context for future review. For example, S8 commented their notes ``\textit{lose a lot of context when reviewing them later on.}''

\subsection{Interview}

To gather more insights about the challenges with digital note-taking and user coping strategies, we invited survey respondents who expressed strong opinions\rnr{---defined as those who selected ``Never,'' ``Rarely,'' ``Frequently,'' or ``Very Frequently'' on taking pictures or recording audio, and those who rated ``Strongly disagree,'' or ``Agree'' (as no participant rated ``Strongly Agree'') on the question ``Note-taking is distracting and makes me miss content''---}to a follow-up \rnr{semi-structured} interview. \rnr{Participants were asked to share their digital note-taking workflow accompanied by their note samples and elaborate on why they missed content, their strategy to catch up, and their opinions and use cases for using pictures and audio recordings to supplement their notes. Each interview took around one hour to complete and was recorded and transcribed to text. One researcher utilized content analysis and affinity diagramming~\cite{luceroUsingAffinityDiagrams2015} to analyze the interview and count common themes.}

In total, 12 survey respondents participated (P1-P12) in the interview \rnr{(6 women, 5 men, 1 non-binary)}, including ten undergraduate students and two graduate students. Eight students were from Engineering and Technology, two from Natural Sciences, one from Math and Statistics, and one from Social Sciences. Seven participants frequently write or draw with tablets, five frequently type with laptops, five frequently take pictures and two frequently record audio. The interview took around an hour, and participants were compensated \rnr{CA\$25} after completion. 

Participants reported that they miss content during lectures because the lecturer is moving too fast (5/12), the content is complex to follow (5/12), they are distracted by their note-taking (5/12), or they are distracted by digital devices and mind-wandering (6/12). Participants miss content when the lecturer is ``\textit{talking fast}'' (P3) or when ``\textit{very important slides came back to back}'' (P4), resulting in participants having ``\textit{to skip a whole concept completely because I am just trying to keep up}'' (P12). Participants also miss content if they struggle to ``\textit{understand complicated graphs}'' (P9). Some participants reported note-taking led to missing content. For example, P6 commented, ``\textit{Taking more notes is getting me way more lost in the lecture.}'' Digital distractions such as phone usage (P3), ``\textit{mind-wondering}'' during less interesting topics (P9), and ``\textit{dozing off}'' (P12) also lead to participants missing content.

Most participants catch up on missed lecture content by marking and jotting keywords (8/12) or asking classmates sitting next to them (6/12), while some participants prefer to focus on live content and ignore taking notes (3/12). For example, participants commented they quickly leave a mark or write down their questions when missing content: ``\textit{I will just try to make a mark in the slides or just my notes where I got lost}'' (P6). Sometimes, participants ask their neighbors (P7) or ask for other students' notes (P3). Furthermore, some participants prioritize their focus on live lectures even if they could potentially miss content ``\textit{because I would not know whether the next topic is important}'' (P4).

Taking pictures and recording audio also help students keep up with lectures. Participants often take pictures as a quick way to capture lecture content (6/12), especially when the lecturer is too fast (3/12) or when they feel overwhelmed (2/12). For example, P6 said, ``\textit{If I don't have enough time, I will take the picture,}'' and P10 mentioned taking pictures when ``\textit{the professor is going much faster than I can write.}'' Five participants noted that pictures provide context to their notes and help index the lecture, even when slide decks are available, because they ``\textit{do not want to sort through the slides again}'' (P8). However, challenges include cumbersome cross-device synchronization (3/12), difficulty organizing pictures (3/12), and concerns about disrupting the class (3/12). For instance, P12 said, ``\textit{Take a picture and then AirDrop it and then copy it. That takes some time.}'' P8 mentioned the pictures ``\textit{just rot and die in my Google Photos.}'' Some participants refrain from taking pictures because it might be ``\textit{annoying to the instructor}'' (P1) and that ``\textit{if I hold on my iPad, it is going to block people's view}'' (P11).

Many commercial note-taking applications allow users to add audio recordings. Moreover, apps such as Notability\footnote{Notability: \url{https://www.notability.com/}} and GoodNotes\footnote{GoodNotes: \url{https://www.goodnotes.com/}} provide playback features that replay handwriting strokes with the recording. However, most of the students in our survey and from previous survey results \cite{borhaniSurveyAnnotationsExtended2024} opt not to record audio. The two participants who frequently recorded audio said it helped them avoid missing content, commenting that they ``\textit{like to recordings of it to make sure that I can refer back}'' (P11), and audio recording ensures they ``\textit{do not miss anything important}'' (P12). When asked about the challenges with audio recording, participants mentioned it is difficult to review (7/12) and create privacy concerns (3/12). Participants find the audio recordings contain redundant or irrelevant information. For example, ``\textit{during the time of review, I will not have the time to listen to the audio again}'' (P4), and ``\textit{the professor sometimes goes off the slides about different topics, and it is not relevant}'' (P7). P9 mentioned that ``\textit{we need to get permission from the professor if you want to record the voice.}''

\subsection{Design Goals}

In summary, our survey showed that students primarily take pictures and use tablets to take notes while reporting that taking notes might cause distractions and lead to missing content. The interview further revealed that a common factor for missing content is the lecture pace is faster than the student's note-taking speed. Students manage this by leaving marks in their notes and moving on to the current lecture content. Students also take pictures for speed benefits but find organizing and inserting them in their notes challenging. Although audio recording keeps a comprehensive record, few students use it due to navigation difficulties and information redundancy. Based on these findings, we designed MaRginalia to:

\begin{enumerate}
    \item Support existing digital lecture note-taking practices with a tablet and stylus.
    \item \rnr{Reduce friction in recording and organizing lecture content by automatically creating snapshots of lecture slides and speech transcripts, allowing users to curate content of interest and annotate them in place.}
    \item \rnr{Enhance engagement and facilitate catching up on missed lecture content by enabling note-taking with a view of the lecture environment and providing easy access to lecture histories for students reviewing missed content.}
\end{enumerate}
\section{MaRginalia}

MaRginalia is an integrated note-taking system for in-person lectures (\autoref{fig:teaser}). \rnr{To meet the design goals, MaRginalia incorporates a mixed reality head-mounted display (MR HMD) with a tablet and stylus, leveraging MR's unique sensing and display capabilities. While the user wears the MR HMD, its world-view camera captures images of lecture slides, and its microphone records the lecturer’s voice. This eliminates the need for users to manually raise a tablet to take pictures or set up and transfer files from a separate camera. Furthermore, MR HMD can seamlessly blend virtual content into the physical world, allowing users to follow the lecture, take notes, and review slides without shifting attention between devices, minimizing the split-attention effect.}

When using the system, the user sets the tablet on \rnr{a table} and wears the MR HMD. The user can take handwritten notes directly on the tablet or use a virtual projection of the tablet in MR while seeing the lecture. The system automatically creates snapshots of lecture slides and the lecturer's speech and displays them in MR, enabling users to add captures of the snapshots. The \emph{spatial panels} in MR also provide navigation tools for users to quickly review these captures so that they can catch up on missed information.

\subsection{Note-taking}

\begin{figure}
    \centering
    \includegraphics[width=1\linewidth]{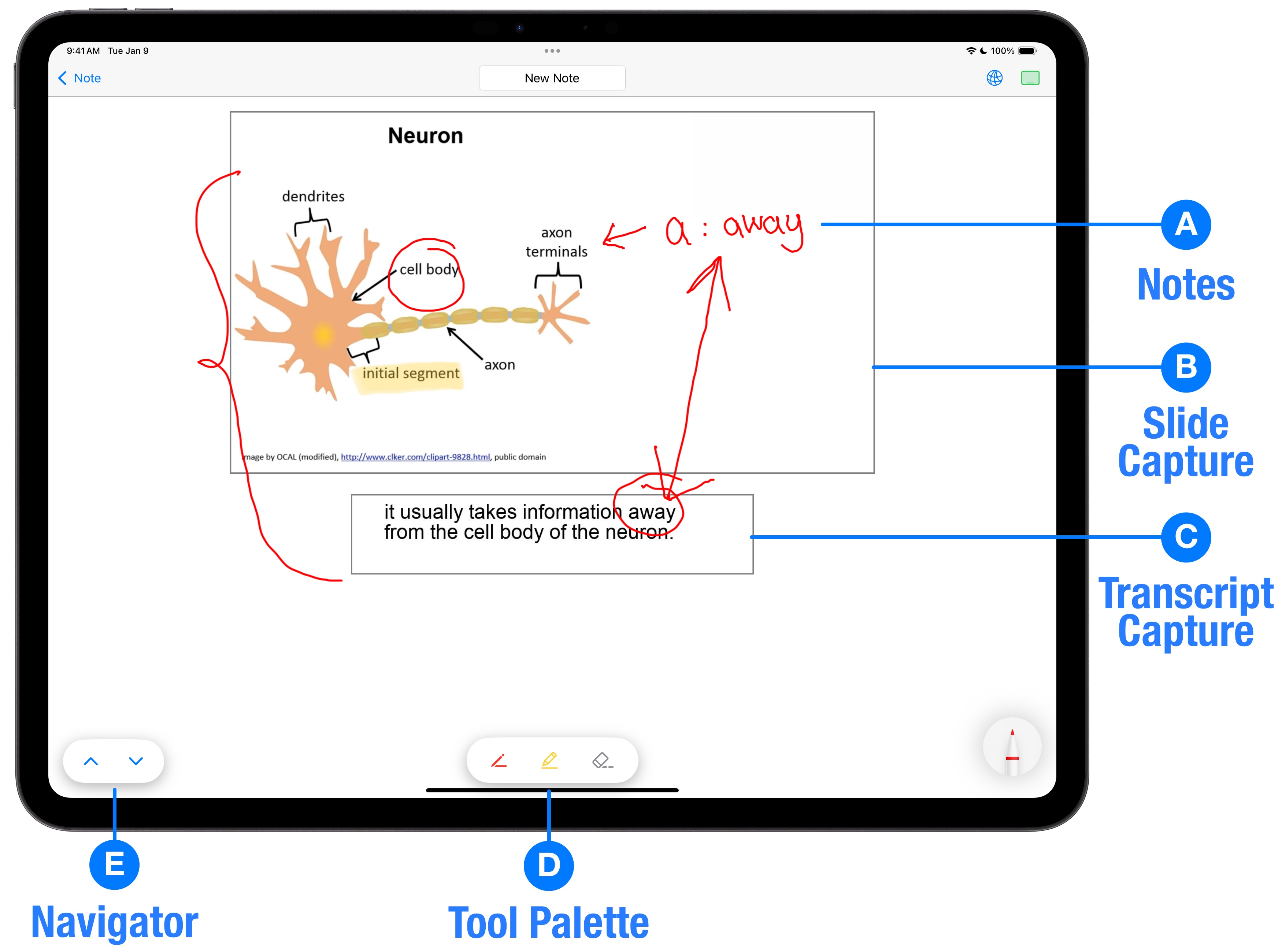}
    \caption{\rnr{Tablet interface. A: User \emph{notes} on the canvas. B-C: User captured snapshots of lecture \emph{slide} (B) and \emph{transcript} (C). D: Drawing \emph{tool palette}, including pen, highlighter, and stroke eraser. E: \emph{Navigator} buttons to scroll between captures.}}
    \label{fig:feat-tablet}
\end{figure}

\autoref{fig:feat-tablet} illustrates MaRginalia's tablet interface with example notes. With MaRginalia, users can take handwritten notes \rnr{on the tablet aligning with their existing note-taking practices.} \rnr{The system only provides three tools, a red pen, a yellow highlighter, and a stroke eraser (\autoref{fig:feat-tablet}D)}, to encourage students to focus on capturing notes rather than on formatting\rnr{, which some survey respondents considered distracting}. The notes canvas is vertically scrollable to provide additional writing space as needed. When the user captures a snapshot of the lecture slide or a block of speech transcription from MR, the capture is chronologically added after the user's note, centered on the drawing canvas, leaving space for users to take notes in the margin. Users can quickly navigate between the captures with the scroll buttons in the \textit{navigator} (\autoref{fig:feat-tablet}E).

\begin{figure*}
    \centering
    \includegraphics[width=1\linewidth]{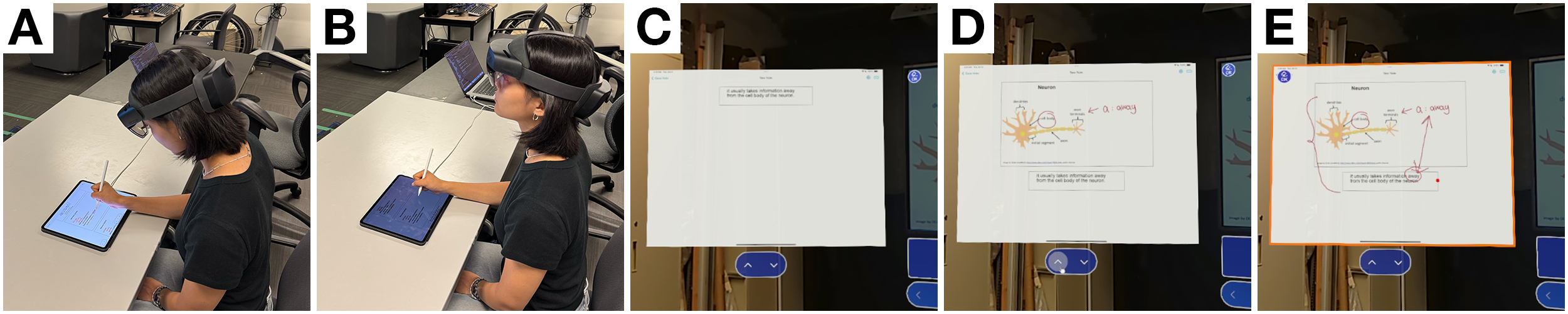}
    \caption{Note-taking on the tablet (A-B) and the spatial \textit{Notes Panel} (C-E). A: Looking at the tablet enables direct handwriting. B: Looking forward, the tablet becomes an indirect input. C: View notes on the Notes Panel. D: Selecting the scroll button navigates between captures. E: Taking notes on the Notes Panel. Notes are synced between the Notes Panel and the tablet.}
    \label{fig:feat-note}
\end{figure*}

\begin{figure*}
    \centering
    \includegraphics[width=0.97\linewidth]{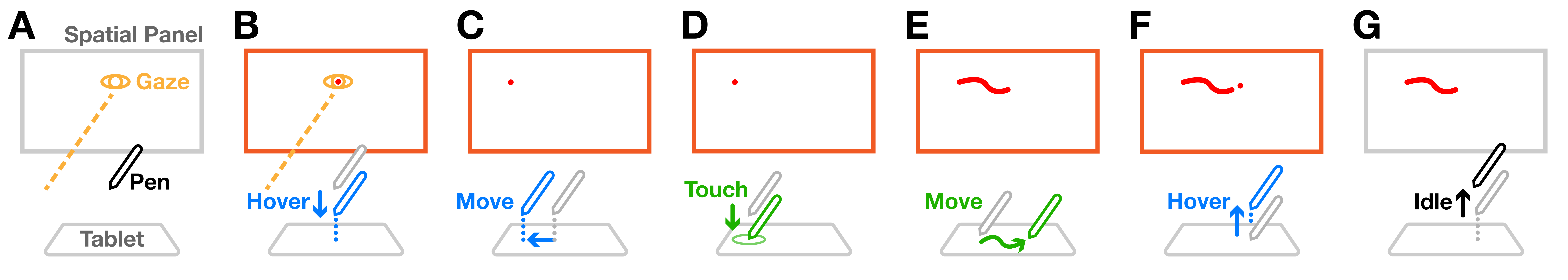}
    \caption{Gaze+Pen creating strokes with spatial panels. A: The pen is idle when far from the tablet. The system continuously detects the user's gaze position. B: Bringing the pen closer to the tablet reveals the cursor. C: Moving the pen while hovering adjusts the cursor position. D: Touching the tablet with the pen engages in an interaction. In this case, begin drawing. E: Moving the pen while touching the tablet draws a stroke. F: Lifting the pen into a hover state completes a stroke and enables cursor position adjustment. G: Lifting the pen resets the system to an idle state and hides the cursor.}
    \label{fig:gaze-pen}
\end{figure*}

MaRginalia also enables users to take notes \rnr{next to the lecture slides in the physical world with the \emph{Notes Panel},} a virtual projection of the tablet display \rnr{in MR} (\autoref{fig:feat-note}C-E). When the user looks forward, the tablet display dims and becomes an indirect input device that allows them to interact with and draw on spatial panels such as the Notes Panel (\autoref{fig:feat-note}A-B). Users can access the scroll buttons and tool palettes next to the spatial panel. The user's notes, drawing tools, and scrolling location in notes are synchronized between the tablet and the Notes Panel, allowing users to take notes on the tablet and in MR based on their preference and workflow.

\subsection{Gaze+Pen Interaction}

\rnr{For users to interact with and take notes on the spatial panels,} we designed a Gaze+Pen interaction technique that combines the user's gaze points in MR and pen input on the tablet for drawing, button selections, and content capturing (\autoref{fig:gaze-pen}). \rnr{Gaze+Pen was inspired by \emph{MAGIC}~\cite{zhaiManualGazeInput1999}, where the fast yet jittery gaze provides the initial cursor position, and then the slow but stable hand provides fine cursor adjustments. }The user operates Gaze+Pen by initially gazing at the area with which they want to interact, then initiating the interaction starting from the gaze point with pen movements. Gaze+Pen operates in three states: \emph{idle}, \emph{hover}, and \emph{pen-down}. In the idle state, the cursor is hidden to avoid distractions (\autoref{fig:gaze-pen}A). The user hovers the pen over the tablet to enter the hover state \rnr{to preview the cursor position}. The system reveals the cursor at the gaze point and highlights the spatial panel to provide visual feedback. The system then decouples gaze from cursor control, allowing the user to adjust the cursor by maintaining hover while moving the pen (\autoref{fig:gaze-pen}C). \rnr{As demonstrated in prior work~\cite{pfeufferGazetouchCombiningGaze2014, bienerBreakingScreenInteraction2020}}, using gaze to reposition the initial cursor position enables interaction beyond the physical size of the tablet, minimizing excessive hand movements across the tablet surface. Decoupling gaze to refine cursor position compensates for potential gaze-tracking errors and prevents erratic cursor shifts from unintentional gaze movements or the user's gaze shifting away from where they wish to interact. Touching the tablet with the pen enters the pen-down state (\autoref{fig:gaze-pen}D) and initiates a pen stroke, which continues as the pen moves across the surface (\autoref{fig:gaze-pen}E). Lifting the pen returns the system to a hover state, showing the cursor at the end of the stroke and allowing further adjustments to start a new stroke (\autoref{fig:gaze-pen}F). Fully lifting the pen from the tablet resets the system to idle (\autoref{fig:gaze-pen}G).

\begin{figure}
    \centering
    \includegraphics[width=0.8\linewidth]{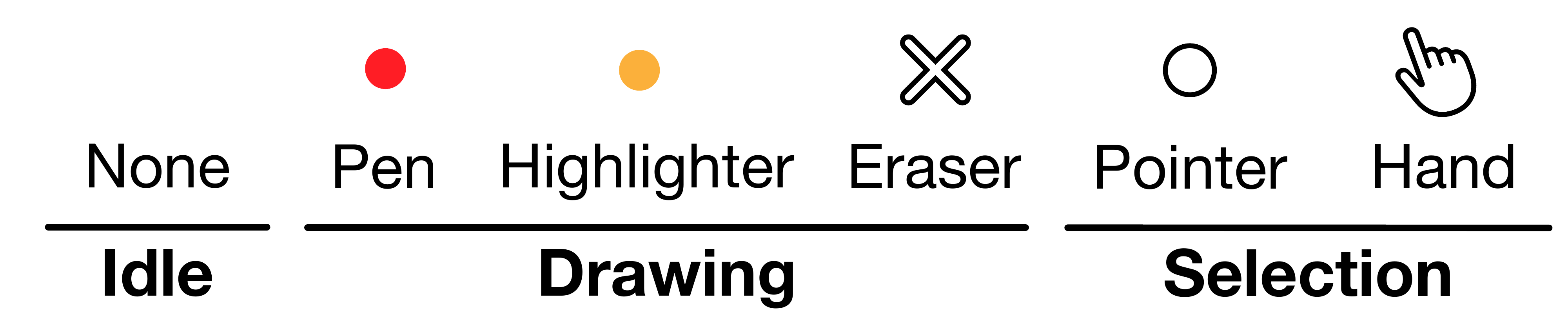}
    \caption{Cursor icons. The cursor hides when idle and dynamically reflects drawing tools and button selection states during hover and pen-down.}
    \label{fig:cursor}
\end{figure}

\begin{figure*}
    \centering
    \includegraphics[width=0.97\linewidth]{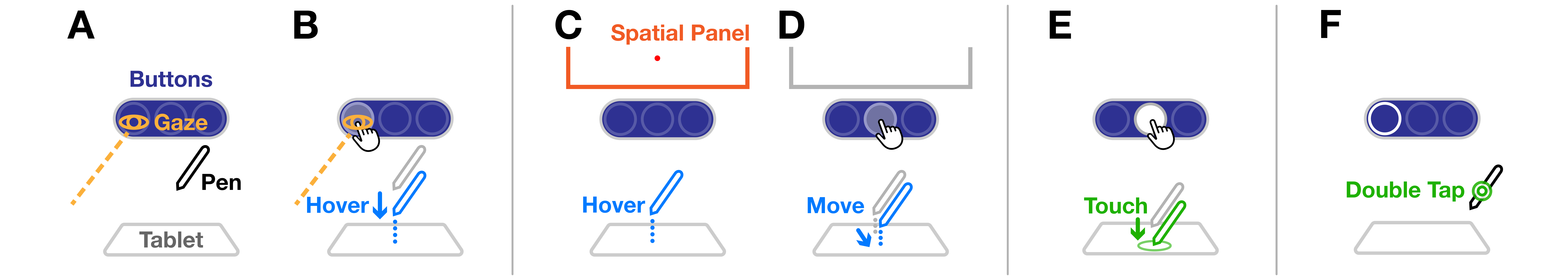}
    \caption{Selecting buttons in MR. A-B: Select the button when the pen is idle by gazing at the button (A) and hovering the pen to reveal the cursor (B). C-D: Move the cursor from a spatial panel (C) to a button (D). E: Tap the pen on the tablet to confirm the button selection. F: Double-tap the side of the pen at any time to switch between drawing tools.}
    \label{fig:button}
\end{figure*}

Buttons control the MR interface, such as scrolling through notes or switching between drawing tools. The Gaze+Pen cursor is attached to the spatial elements' surface and dynamically changes icons to indicate different drawing tools and selection states (\autoref{fig:cursor}). The user selects a button from the idle state by gazing at the button (\autoref{fig:button}A-B) or moving the cursor from a spatial panel to the button while hovering the pen (\autoref{fig:button}C-D). Touching the tablet with the pen completes the selection (\autoref{fig:button}E). The button under the cursor highlights while hovered and flashes white on pen-down, confirming the selection. As the user's gaze shifts the cursor, they can lift the pen, look at the desired button, and tap the tablet for quicker selections. To expedite tool switching while writing, we added a double-tap pen gesture to quickly return to the last-used drawing tool (\autoref{fig:button}F).

\subsection{Capture and Annotate Slides and Transcripts}

\begin{figure}
    \centering
    \includegraphics[width=1\linewidth]{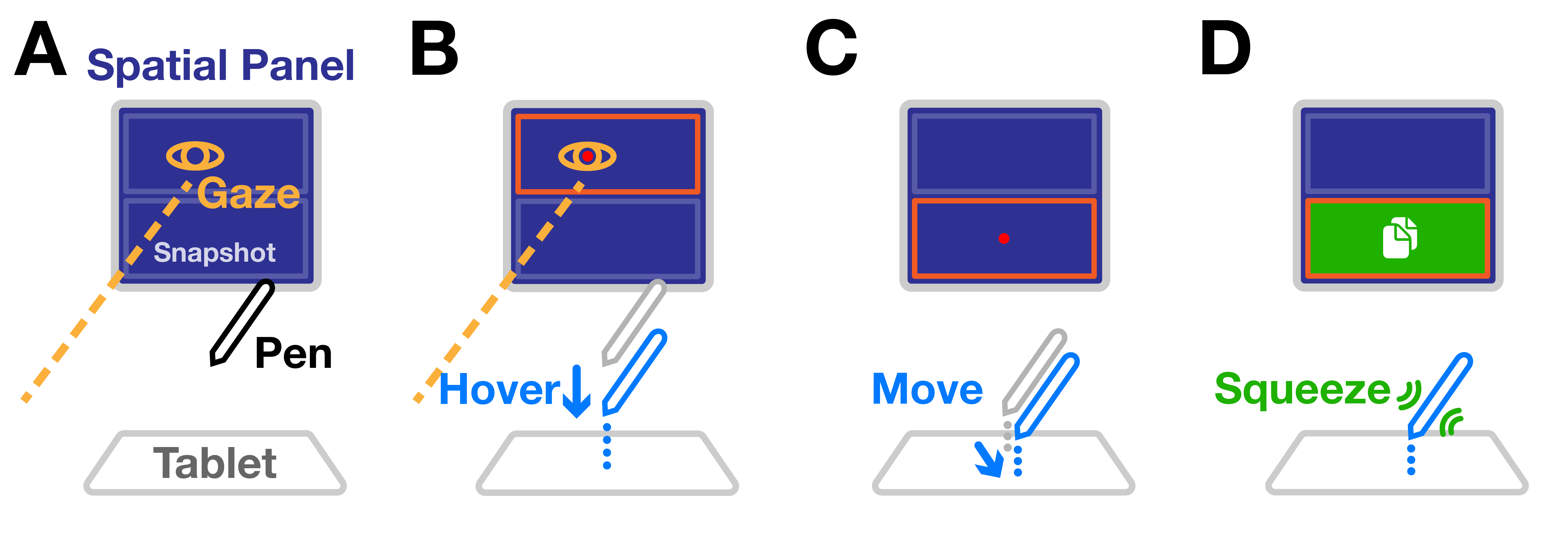}
    \caption{Capturing snapshots with spatial panels. A-B: The user gazes at a snapshot (A) and hovers the pen to reveal the cursor, highlighting the snapshot (B). C: While hovering, the user moves the pen to highlight a different snapshot. D: The user squeezes the pen to capture the highlighted snapshot.}
    \label{fig:copy}
\end{figure}

\rnr{The formative study revealed that taking pictures and recording audio helps students keep up with lectures. MaRginalia automates the process and removes the cross-device synchronization and organization barriers.} MaRginalia updates snapshots of the lecture slides and transcripts, then displays them in the \textit{Slides Panel}'s slide navigator (\autoref{fig:feat-slide}) and \textit{Transcripts Panel} (\autoref{fig:feat-transcript}). The Slides Panel is aligned with the lecture hall display, and the Transcripts Panel is displayed adjacent to the speaker, enabling in-place viewing \rnr{without split attention}. Both panels enable users to capture and annotate \rnr{on the fly, simplifying the review process by eliminating the need to revisit or interpret the entire recording later.} When the user annotates the snapshot, the system automatically creates a capture of the annotated snapshot, adds it to notes, and displays a green check mark to indicate success. Any additional annotations on the same snapshot will be added to the latest capture. The user can also manually capture the snapshots with a pen squeeze gesture. A squeeze on the pen adds the snapshot highlighted under the cursor, accompanied by any annotation, to the user's notes and signals success with a green flash (\autoref{fig:copy}). If the user has existing annotations on the snapshot, the system will clear the annotation in MR (see \autoref{fig:feat-slide}C-D as an example with the Slides Panel).

\subsection{Navigate Lecture History}

\begin{figure}
    \centering
    \includegraphics[width=1\linewidth]{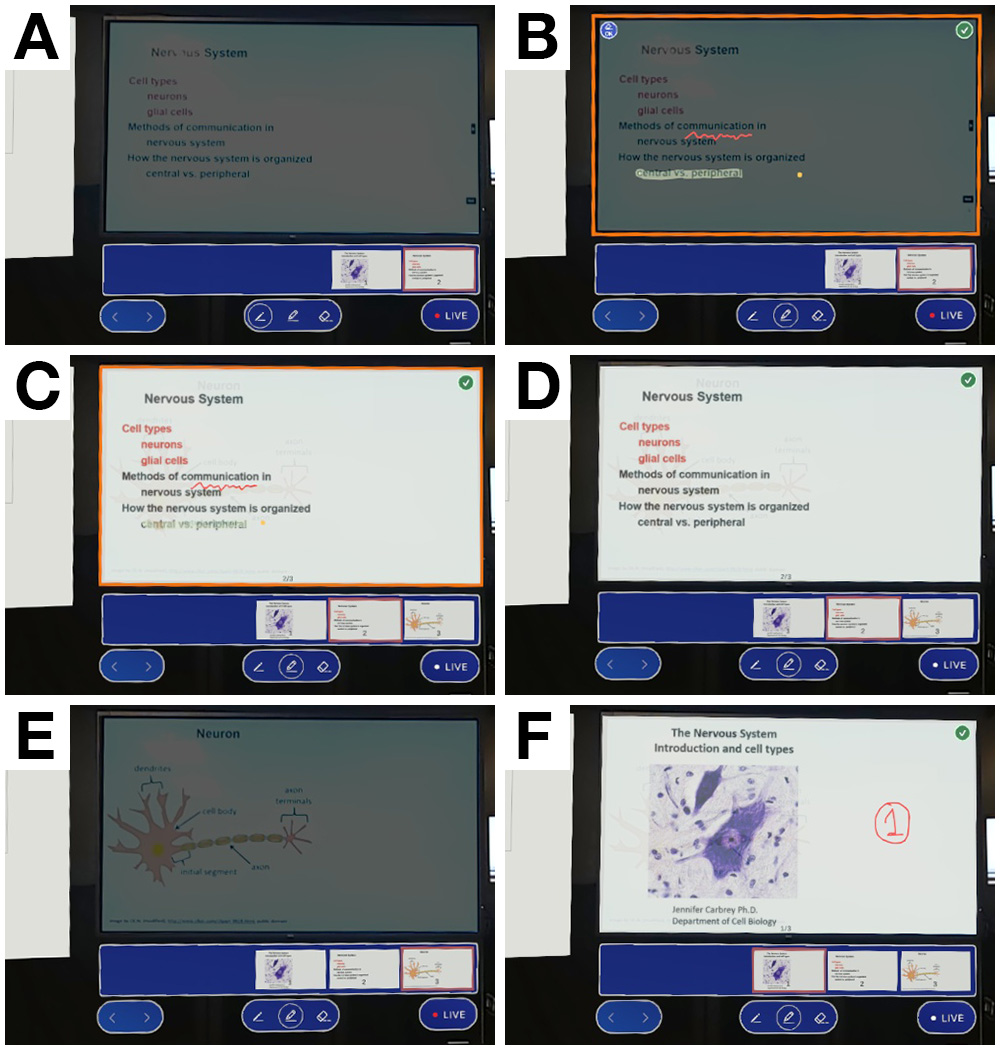}
    \caption{View, annotate, capture, and navigate the \emph{Slides Panel}. A: The panel is aligned with the lecture hall display, showing the latest slide transparently. B: Annotate the current slide. C: The system retains the prior slide when the user continues annotating and the lecturer flips the slide. D: Create a fresh slide capture by squeezing the pen. E: Sync up with the lecture by selecting the live button. F: Navigate to a prior slide by selecting a thumbnail in the slide navigator.}
    \label{fig:feat-slide}
\end{figure}

\rnr{MaRginalia automatically saves previous slides and the lecturer's audio transcript for users to review quickly when they miss content during live lectures.}
Below the current slide, the \emph{Slide Navigator} shows thumbnails of recent slides, with the currently displayed slide highlighted in orange. The system makes the live slide transparent in MR, allowing users to follow the physical screen in the lecture hall if the lecturer is using a pointer. (\autoref{fig:feat-slide}A, B, E). Users can select thumbnails to review any previous slide accompanied by their annotations (\autoref{fig:feat-slide}E-F). The previous slide replaces the current slide overlaid in the lecture hall display, and the \textit{live button} turns from red to white to indicate the user is out of sync (\autoref{fig:feat-slide}B-C and E-F). During live lectures, users might not finish note-taking before the lecturer changes slides. In such cases, MaRginalia overlays the snapshot of previous slides for the user to complete the annotation (\autoref{fig:feat-slide}B-C). To return to the live session, the user can click the live button or select the latest slide thumbnail (\autoref{fig:feat-slide}D-E).

\begin{figure}
    \centering
    \includegraphics[width=1\linewidth]{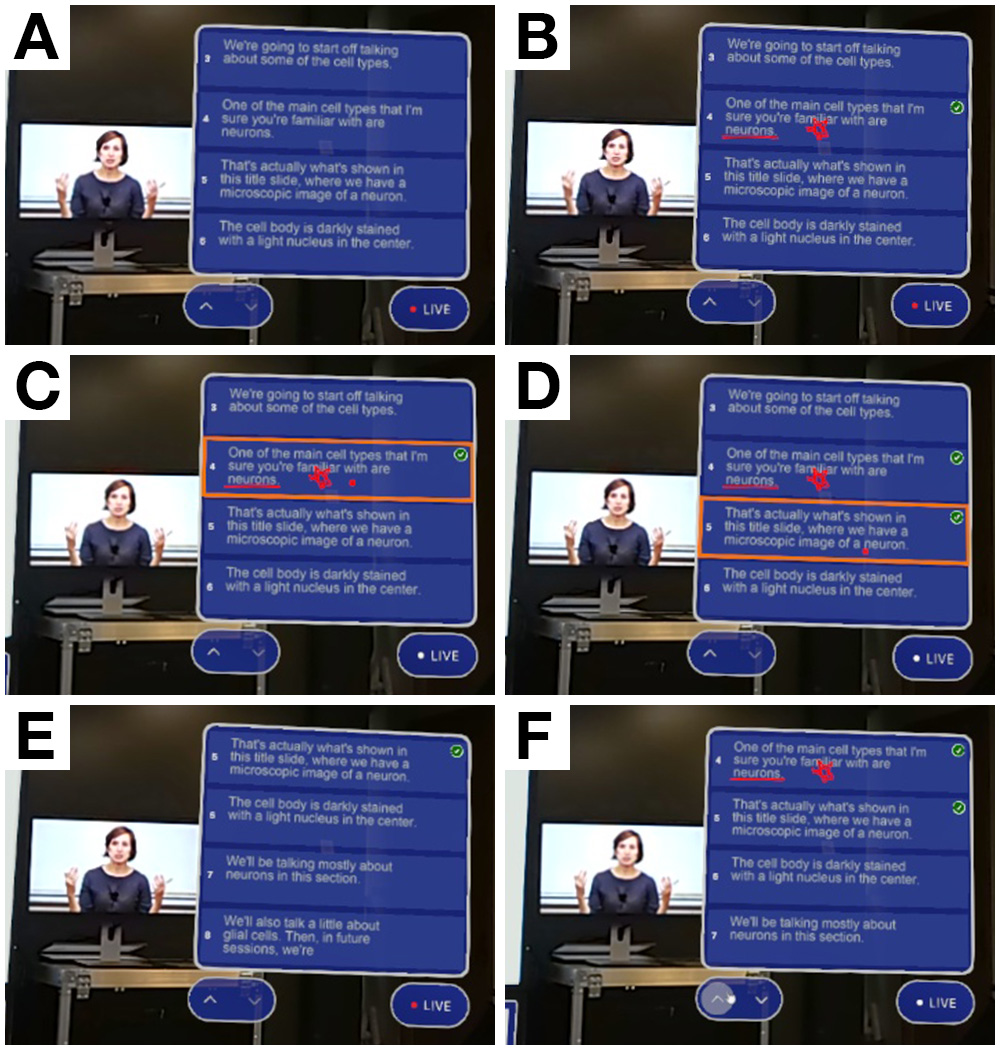}
    \caption{View, annotate, capture, and navigate the \emph{Transcripts Panel}. A: View the latest transcription of the lecturer’s words. B: Annotate a transcript block. C: The system pauses snapshot updates as the user completes annotating. D: Capture the transcript by squeezing the pen. E: Sync up with the lecturer by selecting the live button. F: Navigate to the previous transcript by selecting the scroll button.}
    \label{fig:feat-transcript}
\end{figure}

\rnr{Participants from the interview reported missing fast-paced and complex content. MaRginalia provides transcripts of the lectures' words to facilitate user understanding.} 
The transcript snapshot is updated to display at the bottom of the Transcripts Panel, allowing users to read naturally between the prior snapshots in top-to-bottom reading order (\autoref{fig:feat-transcript}). The scroll button to navigate between the transcript snapshots (\autoref{fig:feat-transcript}E-F). As with the Slides Panel, the live button indicates when the user is out of sync (\autoref{fig:feat-transcript}B-C and E-F), and the system pauses transcript updates during user annotations (\autoref{fig:feat-transcript}B-C). The user can select the live button to return to the latest transcript (\autoref{fig:feat-transcript}D-E). The Slides and Transcripts Panels are paused independently, allowing the user to stay engaged with the live lecture by peaking the slides while reviewing the transcripts and vice versa.

\subsection{Implementation}

\begin{figure}
    \centering
    \includegraphics[width=0.8\linewidth]{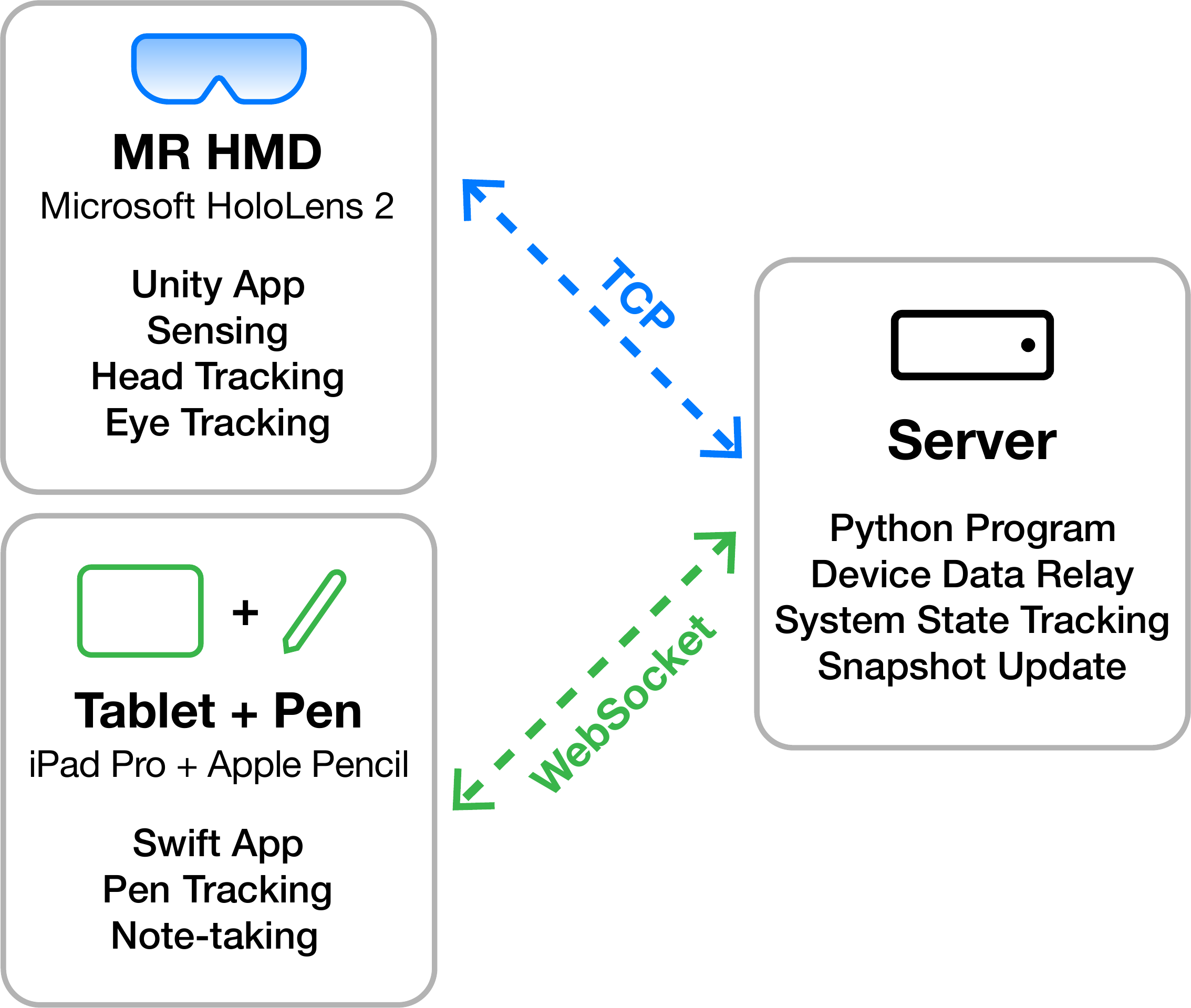}
    \caption{MaRginalia system implementation.}
    \label{fig:system}
\end{figure}

\autoref{fig:system} illustrates the three components of MaRginalia:
1) an MR HMD that displays spatial elements, handles Gaze+Pen interactions, and tracks users' head position; 
2) a tablet and pen for tablet note-taking and pen input; and
3) a server to relay data between devices, track system states, and update the snapshots for the HMD.

In our implementation, for the HMD, we used Microsoft HoloLens 2\footnote{HoloLens 2: \url{https://www.microsoft.com/en-us/hololens/hardware}} for its accessible eye-tracking data. 
We developed a Unity application and used MRTK\footnote{Mixed Reality Toolkit 2: \url{https://github.com/microsoft/MixedRealityToolkit-Unity}} to access gaze data for interaction at 30Hz and the \emph{hl2ss} plugin~\cite{dibeneHoloLensSensorStreaming} to stream data between the HMD and the server over TCP. 
For the tablet and pen, we used an Apple iPad Pro\footnote{iPad Pro: \url{https://www.apple.com/ipad-pro/specs/}} and Pencil Pro\footnote{Pencil Pro: \url{https://support.apple.com/en-us/120123}} for their hover and gesture detection capabilities.
We developed a native Swift application that registers the pen input at 120Hz and streams data between the tablet and the server via WebSocket. We used PencilKit\footnote{PencilKit: \url{https://developer.apple.com/documentation/pencilkit}} to render and store notes on the tablet.
As sensing is not the focus of this prototype, the current prototype preloads the lecture slides and transcripts as images on the server. Time events trigger the server to update the HMD with slides and transcripts at runtime. As sensing capabilities improve, the system can create snapshots at runtime, eliminating the need to preprocess the lecture material.

\section{Evaluation}

To explore the perceived utility and usability of MaRginalia, we conducted a study where students took notes using the system in a simulated in-person lecture environment. Video recordings of lectures were played with \rnr{the lecture slides and a talking head of the lecturer} visible on two monitors (\autoref{fig:study-environment}), and the system updated preloaded snapshots of the slides and transcripts in spatial panels based on time events on the server.

\subsection{Task}

\begin{figure}
    \centering
    \includegraphics[width=1\linewidth]{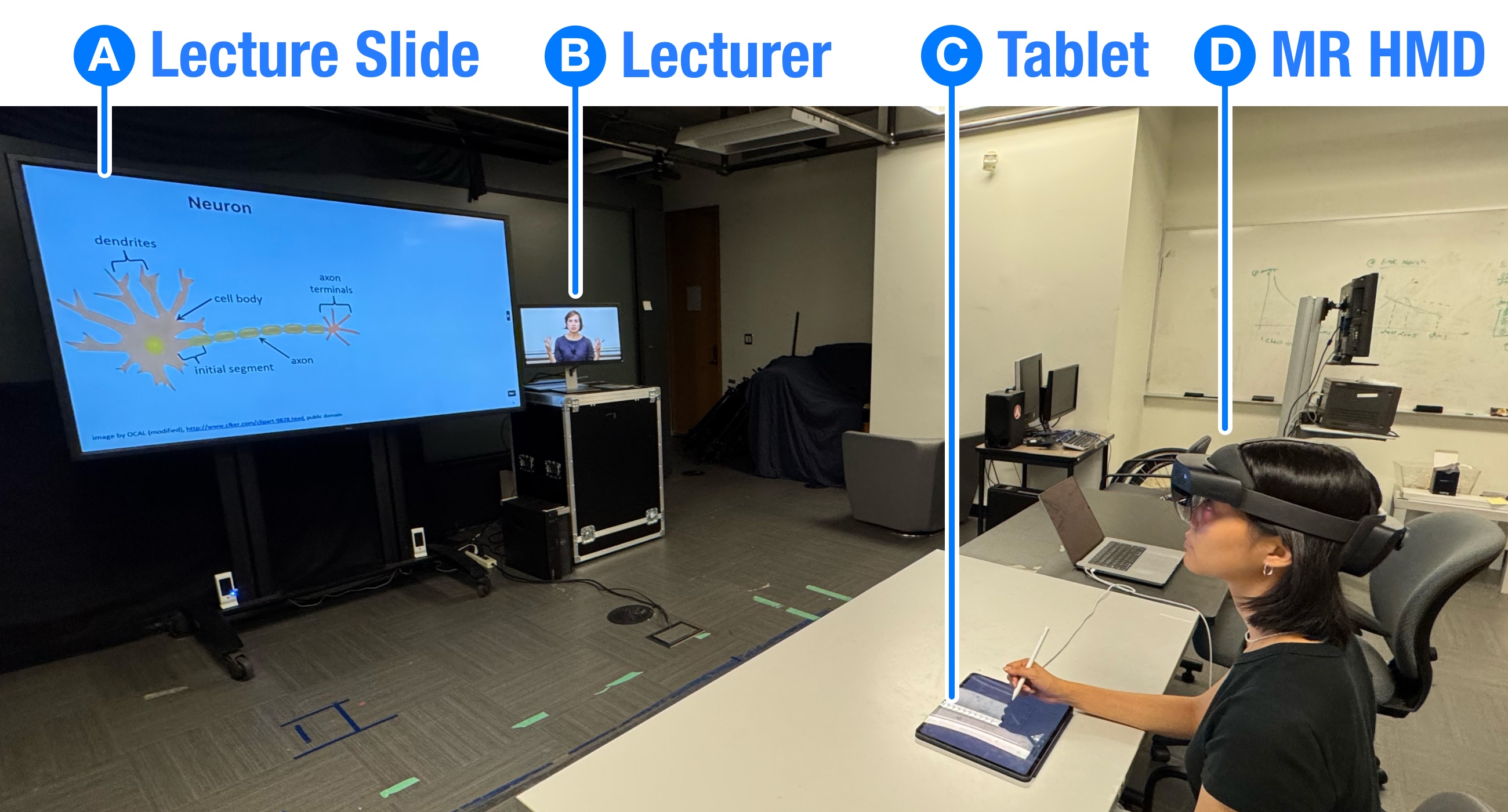}
    \caption{\rnr{User study environment simulating an in-person lecture. A: Display of the \textbf{lecture slides}. B: Display of the \textbf{lecturer}'s face and upper body. C-D: The \textbf{tablet} and mixed reality head-mounted display (\textbf{MR HMD}).}}
    \label{fig:study-environment}
\end{figure}

For the main task, we instructed participants to take notes while watching the lecture video (\autoref{fig:study-environment}A) as students learn the topic for the first time and use the system how they imagine themselves using it. To accommodate differences in participants' backgrounds and interests in the topic, we selected two publicly available lectures for the users to choose from before they began the task: a biology lecture\footnote{Respiratory System Lecture: \url{https://www.coursera.org/lecture/physiology/anatomy-and-mechanics-PZUoc}} and a history lecture\footnote{Michael Faraday Lecture: \url{https://www.youtube.com/watch?v=OBHUrAFLQW8}}. We trimmed the lectures to 12 minutes and matched the information density between the two lectures. We edited videos to display the lecturer on a separate monitor of the slides (\autoref{fig:study-environment}B). We pre-processed the video to create timestamped snapshots of the lecture slide per presentation build and transcripts of the lecturer's speech. 

\subsection{Participants}

Twelve students \rnr{(7 women and 5 men)} participated in the study, including four undergraduate students in their third year or beyond, one Master's student, and seven PhD students. Eleven participants studied Engineering and Computer Science, and one studied Biology. None of the participants frequently or very frequently use MR devices (6 Never, 3 Rarely, and 3 Occasionally), eye-tracking devices (7 Never, 3 Rarely, and 2 Occasionally), or indirect drawing tablets (5 Never, 6 Rarely, and 1 Occasionally). More participants have experience with tablets such as iPads (4 Very Frequently, 4 Frequently, 4 Occasionally, and 1 Never). All participants regularly take digital notes during lectures, with nine writing on tables, eight drawing on tablets, ten typing with a keyboard, nine taking pictures, and none recording audio. Seven participants selected the Biology lecture, and five selected the History lecture. Participants reported an average interest of 3.92/5 on the lecture subject and 2.67/5 on material familiarity.

\subsection{Procedure}

After collecting participants' consent and basic demographic information, we introduced the system through a pre-recorded tutorial video. At any point, participants were allowed to ask for clarifications. We asked participants to wear the HMD and complete its eye-tracker calibration. We then launched the system for the participants. Before the main task, we provided a 10-minute step-by-step tutorial, with the lecture paused for participants to learn Gaze+Pen interaction and system features. We asked them to think aloud, provide initial feedback, and raise any questions. After familiarizing themselves with the system and successfully performing all the features, we asked them to explore it freely with a 5-minute live lecture. This allowed participants to experience the features, practice using them freely, and imagine how they would use the system. After a short break, we asked participants to complete the 12-minute main task. Finally, we asked participants to answer the System Usability Scale and subjective rating questionnaires, followed by a 20-minute semi-structured interview on their workflow \rnr{and perceived usability and usefulness of the system features}. 

We recorded the study environment from an angle similar to that in~\autoref{fig:study-environment} and screen-recorded the tablet interface. The audio in the recordings was transcribed for analysis. However, we did not record the HMD interface, as doing so caused the HoloLens HMD to overheat, leading to glitches and occasional crashes. Instead, we logged participants’ interaction events for further analysis. The study took around 75 minutes to complete. We compensated participants with \rnr{CA\$20} upon study completion.

\subsection{Results}
We analyzed user behavior quantitatively based on their interaction log, produced notes, and questionnaire responses. \rnr{We transcribed the semi-structured interviews, and one researcher conducted a framework analysis, organizing the data into categories aligned with the system’s workflow and features to identify patterns in usability and utility.}

\subsubsection{Quantitative Results}

\begin{figure}
    \centering
    \includegraphics[width=1\linewidth]{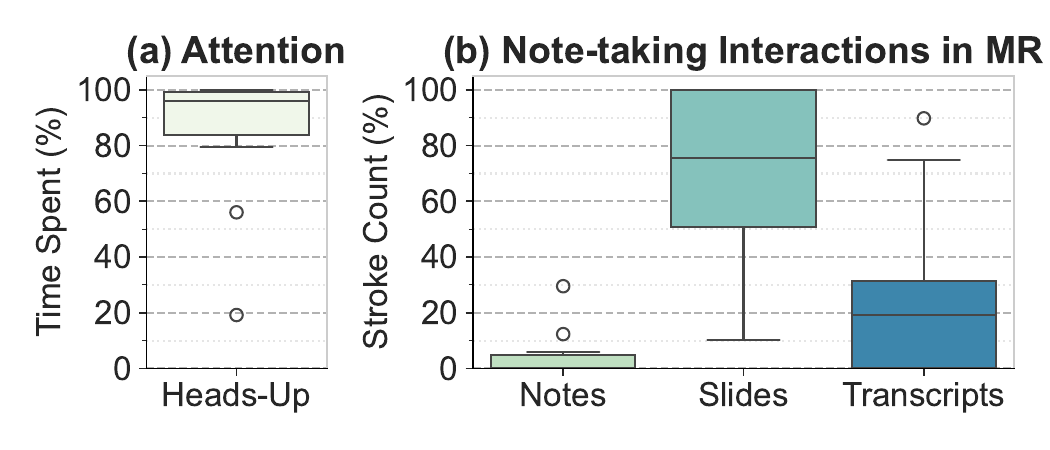}
    \caption{User interaction behavior. (a) Percentage of time spent heads-up in MR per user. (b) Percentage of drawing strokes in the Notes, Slides, and Transcripts Panel.}
    \label{fig:behavior}
\end{figure}

\begin{figure*}
    \centering
    \includegraphics[width=1\linewidth]{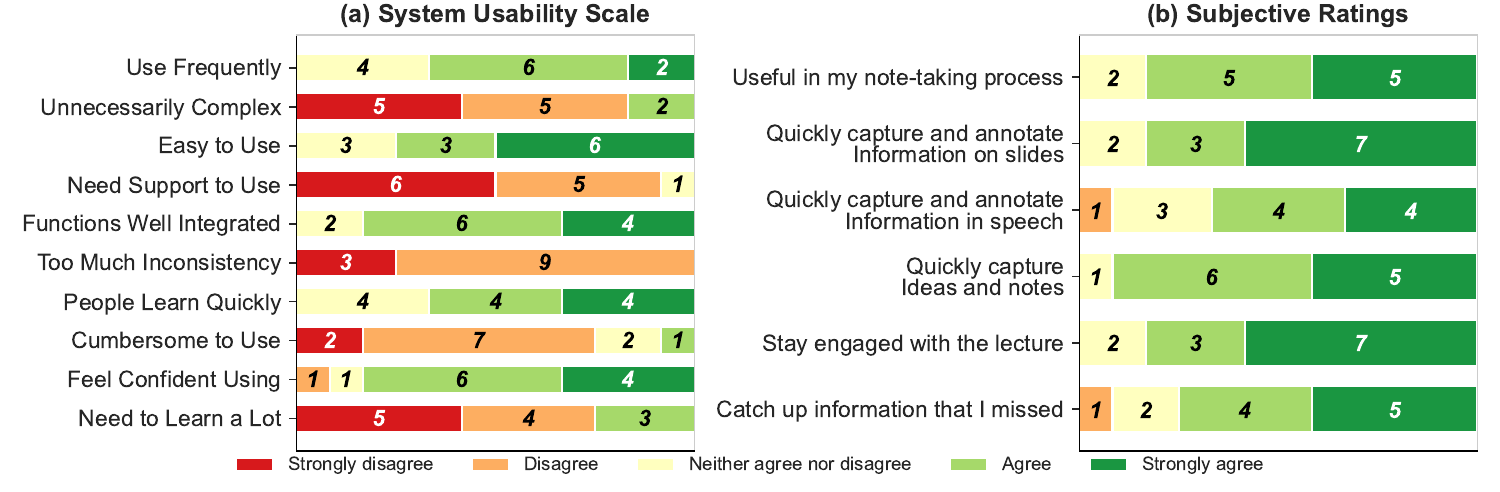}
    \caption{\rnr{Subjective feedback from evaluation study with 12 students. (a) System Usability Scale questions. (b) Participant ratings on the overall usefulness of the system and how the system enables them to accomplish various functions of note-taking.}}
    \label{fig:ratings}
\end{figure*}

All participants captured a total of 103 slide snapshots (median=8.5, IQR=3), and nine participants captured a total of 144 transcript snapshots (median=11, IQR=14), while three participants did not capture any transcript snapshots. Of the snapshots captured, 73.8\% of the slides and 87.5\% of the transcripts were added by MaRginalia automatically when participants annotated on the snapshots, while the rest were manually captured with the pen gesture. \autoref{fig:behavior}a shows the participants' attention behaviors. Participants spent more time in MR than on the tablet, with an average of 85.3\% of time spent heads-up. In this regard, P7 and P12 are outliers who took notes on the tablet and used MR only to capture slides and transcripts. \autoref{fig:behavior}b shows the participants' interaction behaviors in MR. Between the three spatial panels, participants mostly drew on the Slides Panel (all participants), less on the Transcripts Panel (7/12 participants), and significantly less on the Notes Panel (4/12 participants). In total, ten participants used the lecture history navigation feature of the Slides Panel for a total of 52 times (median=3.5, IQR=5), and eight participants used it of the Transcripts Panel for a total of 51 times (median=4.5, IQR=5.5).

\paragraph{MaRginalia achieved a ``good'' adjective rating on the System Usability Scale, with most users agreeing that it effectively met its design goals.} In general, the participants gave positive feedback on the usefulness of MaRginalia. The system received an average System Usability Scale (SUS) score of $77.08\pm14.30$, \rnr{a ``good'' adjective rating in SUS benchmark \cite{bangorEmpiricalEvaluationSystem2008}}. \autoref{fig:ratings}a details participants' ratings on the questions. The participants found the system easy to use and the features well integrated. While three participants identified that the system requires a lot of learning to use, two participants found it unnecessarily complex, and one found it cumbersome to use. \autoref{fig:ratings}b summarizes participants' subjective ratings on MaRginalia's features and functions. Most participants agreed that MaRginalia is helpful for note-taking, allows quick capture of notes and information, and helps them stay engaged and catch up on missed information. \rnr{One participant (P1) found the system not useful for quickly capturing nor annotating speech and catching up on missed information as they reported challenges in reading the slides and listening to the lecturer simultaneously.}

\subsubsection{Qualitative Results}

\paragraph{MaRginalia enables easy and seamless note-taking, while Gaze+Pen interaction is unfamiliar to some.} Participants found the system to be ``\textit{seamless}'' (P7) and ``\textit{a nice integration}'' (P10) that is useful and makes note-taking ``\textit{easier}'' (P6) and ``\textit{convenient}'' (P3). MaRginalia is ``\textit{like an all-in-one spot for note-taking}'' (P5) that helps users ``\textit{select what I needed}'' (P5). When using MaRginalia, the user can ``\textit{put emphasis on certain parts of the slides and capture the transcript of important things}'' (P11). All participants could quickly learn and perform the basic interactions and tasks during the system tutorial in under three attempts. When they first tried the system, some participants commented that the system was ``\textit{smooth}'' (P5), ``\textit{cool}'' (P4, P6), ``\textit{intuitive}'' (P10), and ``\textit{makes sense}'' (P6, P8, P11). However, although all participants could complete the tasks, some commented that the indirect Gaze+Pen interaction for writing is unfamiliar, requiring more time to get used to it (P3, P7, P9, P12).

\begin{figure*}
    \centering
    \includegraphics[width=1\linewidth]{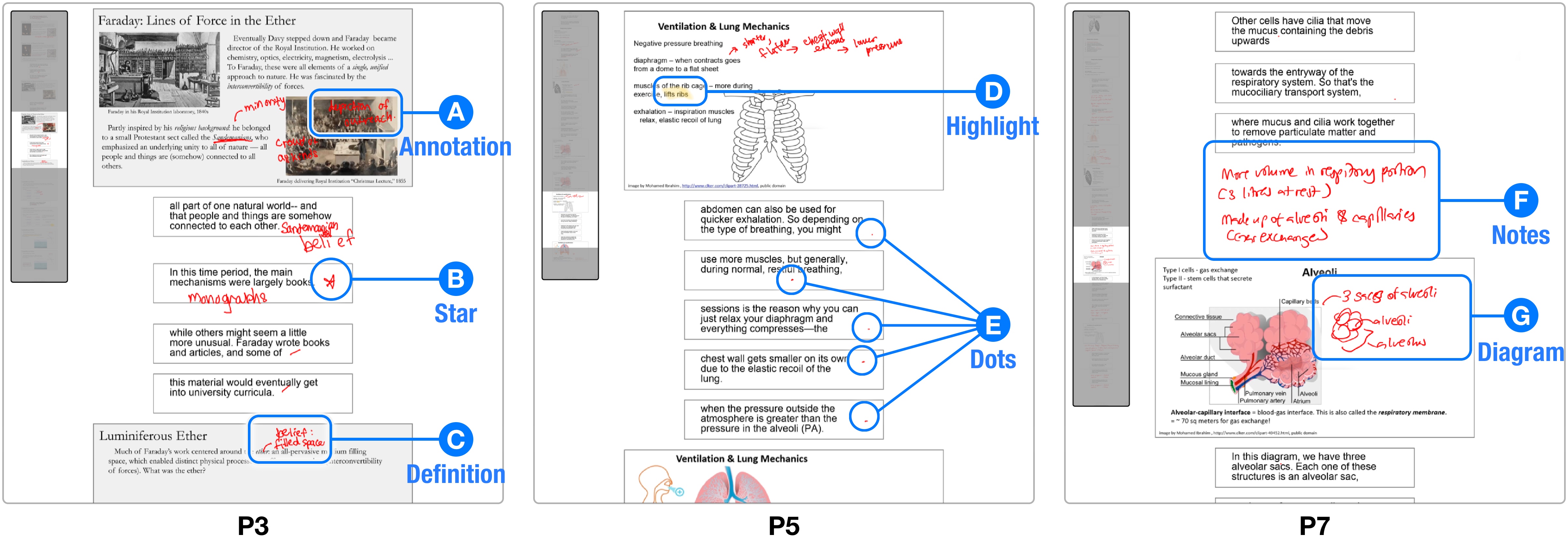}
    \caption{\rnr{Excerpts of participants' notes. P3 added \emph{annotation} over slide image (A), drew a \emph{star} on the transcript to highlight the lecturer's speech (B), and wrote a short \emph{definition} of a term on slide (C). P5 used a marker to \emph{highlight} text (D) and added \emph{dots} on the transcript to quickly capture the lecturer's speech (E). P7 added \emph{notes} between the captures (F) and drew a \emph{diagram} to annotate slides (G). The minimap visualizes the excerpt's location in participants' notes and is not part of the system.}}
    \label{fig:notes}
\end{figure*}

\paragraph{MaRginalia is flexible to support different workflows.}
\autoref{fig:notes} shows excerpts from the notes taken by three participants using MaRginalia. Each participant adopted a distinct workflow. P3 took the history lecture with text and images. They underlined key phrases in the lecturer's speech (\autoref{fig:notes}A), emphasized some key points with a star symbol (\autoref{fig:notes}B), and annotated slides with clarifications of terms and annotated slides with definitions of terms and descriptions of images (\autoref{fig:notes}C), all using the spatial panels. P5 and P7 took the biology lecture with a mix of text, figures, and bullet points. P5 used the squeeze gesture to capture slides, then added annotations on the Slides Panel. For quick successions of multiple transcripts, however, P5 primarily captured them by adding a dot on the Transcripts Panel (\autoref{fig:notes}E), citing it was ``\textit{easier to tap than to squeeze.}'' P7 only used the squeeze gestures to capture slides and transcripts from the spatial panels. They then annotated the captures with additional notes in the margin on the tablet, as they found direct writing more ``\textit{natural}'' than Gaze+Pen on spatial panels. \autoref{fig:notes}F shows the system inserting captures after the users' handwriting to avoid overlapping. P7 commended the system that ``\textit{with this, you just click the pen, and it adds (captures) to your notes,}'' and one could ``\textit{focus on writing notes.}''

\paragraph{Slides Panel allows users to quickly capture slides and conveniently annotate them.} The participants found capturing slides very useful and commended the features being ``\textit{convenient}'' and ``\textit{fast}'' (P3, P7). P1 mentioned that they usually ``\textit{never have enough time to take a picture, transfer it to my laptop and scribble over it}'' when lecturers do not provide slides before lectures. In addition, MaRginalia helped participants avoid the ``\textit{need to download and copy the slides, which is cumbersome}'' (P11) even when slides are provided. The feature of annotating on slides is also consistent with participants' existing note-taking methods. The participants said that annotating on the Slides Panel is natural, as ``\textit{it's natural to just look at the screen}'' (P2), direct: ``\textit{As soon as I see it, I can annotate directly}'' (P4), and convenient, as the system ``\textit{makes it even easier and more convenient to use,}'' without the need to ``\textit{manually scroll up and down to find the history (slide to annotate)}'' (P3).

Participants proposed features to improve the annotation experiences with the Slides Panel. Four participants suggested providing more pen color options, and two suggested the system automatically straightens underlines. P5 and P12 wished that if the lecturer wrote and drew on the slides, the system could recognize and automatically add those annotations to the user's notes. P8 wished the system could treat the different builds of the same slide, such as when bullet points are revealed line by line as part of the same slide.
This way, their annotations would stay visible as the lecturer progressively reveals the material.

\paragraph{Transcripts Panel helps users follow the lecture and quickly capture key points.} Participants found the Transcripts Panel useful as it supported them to ``\textit{select or pick from what the lecturer said}'' (P2) and ``\textit{do not have to rewrite what the lecturer is saying}'' (P3). The participants used the feature to follow along (P3, P5) and look up the spelling of an unfamiliar word (P6). Capturing and annotating a transcript helps participants keep up with the lecture if they write slower than the lecturer talking (P12) or miss writing down the information they heard (P6, P9). The participants also mentioned that capturing transcripts felt similar to how they usually transcribe the lecturer's words by hand (P2). Capturing transcripts to their notes also allows participants to quickly fill in the gaps with definitions of new terminologies (P4) and acronyms (P5).

Some other participants found the transcripts less useful during live lectures because they feared missing information (P8, P9) and ``\textit{felt like I was working in the past}'' (P1). These participants imagined the transcript to be more useful when reviewing the notes. P3, P6, and P10 wanted the transcript captured to the notes to be better associated with the slides for easier review. P7 wanted the ability to reorganize transcript captures in notes as they found themselves capturing the transcript out of order. P3 wanted the system to highlight definitions and keywords in the transcript for easy identification when navigating. P2 wanted the Transcripts Panel to be placed below the slides so that they could avoid the need to turn their heads to see all transcripts.

\paragraph{Notes Panel allows users to preview notes, take advantage of margin space, and connect between captures.} Participants found the Notes Panel helpful for previewing their captures and annotations without looking down at their devices and being distracted (P2, P10). It was also useful for writing notes, as it provided extra space to write in the margin (P3, P6) and for associating slides and transcripts (P1) and ``\textit{connect the dots}'' (P5). Other participants found the feature less useful as all features are available on the tablet (P4, P12), or they worried looking away from the lecture to the Notes Panel would be distracting (P8). To make the Notes Panel more useful, P4 and P6 wanted the ability to zoom in and write smaller text, and P8 proposed to integrate the panel with the Slides Panel so that they only need to focus on one spatial panel at all times.

\paragraph{Note-taking with the tablet due to habits and familiarity.} Participants found looking down at the tablet to write longer sentences to be ``\textit{direct}'' (P1), ``\textit{familiar}'' (P9), and ``\textit{natural}'' (P7), mentioning that ``\textit{it is just how I normally take notes}'' (P10). P4 said they liked using the tablet because it felt physically closer to them. Participants who mainly used the tablet to annotate slides and take notes found writing on the tablet to be more natural (P7, P12) but also commented that ``\textit{if I get more used to it (Gaze+Pen interaction), I will probably not look down as much because I feel like looking down on the iPad does risk missing a slide}'' (P12).

\paragraph{Users stay engaged with the lecture because they see the lecture heads-up and can notice page turns.} Participants found that MaRginalia helped them stay engaged during the lecture. Since the user can look at the lecture while annotating and looking forward, participants found the system allows them to ``\textit{keep looking at the material}'' (P1) and be more focused because ``\textit{you are always seeing the sides and information}'' (P9) and ``\textit{limiting distractions}'' (P4) with all the necessary features accessible with spatial panels. The participants also found that by looking forward, they could ``\textit{look at the professor's body language}'' (P4) and ``\textit{stay focused looking at the professor while looking at the slides}'' (P10). As the system automatically updates slides as the lecturer changes slides and provides visual cues, the participants found the system helpful in keeping track of the lecture so that they could ``\textit{easily go back and start adding more detail to some of my notes while being able to see like when she (the lecturer) moves on to a more important topic}'' (P8).

\paragraph{MaRginalia enables users to catch up on information missed by lecture history.} Participants found MaRginalia helpful for catching up as it keeps snapshots of previous slides and the lecturer's speech so that they could navigate to review ``\textit{if I forget to save slides}'' (P5) or ``\textit{when the professor is going fast}'' (P3). Participants found that they could ``\textit{easily reread back on what happened previously}'' (P3) and ``\textit{could easily add that to my notes and move on to the next thing that she (the lecturer) was saying}'' (P6). Some participants found the transcript particularly useful for catching up on missed information. P2 commented that they ``\textit{know exactly what the lecturer said,}'' whereas the ``\textit{slide does not really give me that information,}'' and P3 commented that the transcript was ``\textit{digestible to read.}'' Other participants found listening to live lectures while reading previous transcripts challenging and commented, ``\textit{I was unable to focus on the lecture and whatever I was reading}'' (P12), and the action worried them that they ``\textit{missed on new stuff}'' (P1). To address this, P12 suggested syncing the transcript and its associated slide when quickly reviewing for an easier catch-up experience.

\paragraph{System usability.} MaRginalia received positive feedback from the participants, who commended it as ``\textit{very easy to learn}'' (P12), ``\textit{feels intuitive}'' (P6), and ``\textit{very easy to use and easy to understand}'' (P8). P6 enjoyed the layout integration of notes, slides, and transcripts, allowing them to navigate the interface better and ``\textit{work with them at the same time.}'' P3 and P5 enjoyed the Gaze+Pen interactions as it enabled them to move pen strokes and select buttons quickly.

Participants who perceived the system as less usable encountered challenges with indirect Gaze+Pen interaction. P1, P2, and P8 struggled to hover the pen without looking at it, often inadvertently lifting it to an idle state. This caused the cursor to follow their gaze, shifting to a position different from where they expected the next pen stroke with direct input. P12 noted that the system’s switch between direct and indirect input when moving their head forward made it difficult to write on the tablet while glancing at slides. P5 and P8 reported that the pen occasionally moved outside the tablet during indirect interaction and suggested adding MR visuals to alert users when the pen becomes close to the tablet’s edge.

Although helpful, some participants found the system cumbersome to use. P4 commented that the system requires one to wear an HMD and use a tablet, which is more burdensome than just bringing a tablet to lectures. P1, P2, and P11 reported that they often moved their heads horizontally to fully see the different spatial panels because HMD's display has a small field of view. P1 and P2 were overwhelmed by trying to pay attention to all spatial panels. They suggested the system could provide customization and limit functionalities to focus on the ones they use.
\section{Discussion}

In this paper, we contribute MaRginalia, an MR and tablet note-taking system that allows users to take notes in MR while remaining engaged with the lecture and using existing tablet and pen note-taking practices. MaRginalia supports creating snapshots of the lecture slides and visualizing the lecturer's speech with transcripts so that users can quickly capture lecture content into their notes, annotate over the captures in place, and quickly review the lecture history if they missed content. To support drawing and interacting with the MR spatial elements with a tablet and pen, we propose a Gaze+Pen interaction technique that enables fast cursor repositioning through gaze and indirect input via pen.

The results of our study showed that participants found MaRginalia's support of the existing tablet and pen note-taking paradigm while extending it to MR and providing additional features useful for note-taking. Furthermore, participants exhibited different note-taking workflows ranging from primarily using the tablet, equally using features on the tablet and in MR, and solely annotating on slides with spatial panels. This suggests that MaRginalia is flexible in supporting participants' note-taking preferences. Participants remarked that MaRginalia's content capture and annotation features were similar to their existing note-taking practices but made the process faster and more convenient. The annotation and note-taking features on spatial panels allow participants to see the lecture as it proceeds and stay engaged without unknowingly missing out on content while taking notes. Reviewing previous slides and speech transcript history also helps students catch up on missed information. With limited prior experience with MR and indirect gaze and pen input technique, all participants could use all features successfully after a short tutorial. The system also received a \rnr{``good''} rating on the System Usability Scale, validating the design.

MaRginalia is the first system to explore MR note-taking for lecture settings. \rnr{It leverages MR's unique display capability to seamlessly blend digital content into the physical world and anchor spatial panels in the lecture hall.} Users can access all the note-taking features via a Gaze+Pen interaction, \rnr{which enables users to annotate lecture content and take notes without needing to look down at their devices, thus reducing the split-attention effect~\cite{swellerSplitAttentionEffect2011}. During the study, we observed participants predominately took long handwritten notes on the tablet rather than using the spatial Notes Panel. We speculate two reasons:} First, HoloLens' limited field of view \rnr{discouraged users from interacting with the Notes Panel, as they had to rotate their heads to view the Notes Panel in full}. This issue could easily be resolved through increased HMD field of view and adaptive UI placements. Second, although the participants found the Gaze+Pen interaction natural and easy to use, getting used to indirect drawing input requires time. With less than an hour to learn and use the system, some participants preferred using the tablet due to existing habits and unfamiliarity with indirect drawing interaction. Participants recognized the benefits of writing notes in spatial panels and expressed interest in taking long handwritten notes with the MR Notes Panel if given more time to familiarize themselves with the interactions.

\rnr{MaRginalia also utilizes MR HMDs' world-view cameras and microphones to enable users to capture and annotate lecture slides and lecturer's words quickly. Photographing slides and recording audio on a separate device can be cumbersome to incorporate into notes and contain redundant information that is overwhelming to review. MaRginalia instead enables users to capture and annotate only the lecture content they are interested in directly in MR. This way, users can focus on the lecture and take notes quickly as the lecture progresses, similar to the benefit of photographing and embedding lecture slides to notes in a VR note-taking system~\cite{chenIVRNoteDesignCreation2019}.} To evaluate the interactions and features, currently, MaRginalia used pre-processed lectures to simulate a high-quality and robust perception system for generating snapshots. This was mainly due to the hardware limitations of the HoloLens, which can be improved as the hardware evolves to make the system more integrated. Future work can explore incorporating dynamic slide content, such as capturing the lecturer's whiteboard writing during live lectures. Finally, with the camera on HMD capturing the lecture hall, privacy concerns need to be addressed through more salient hardware recording indicators to people around the user and techniques limiting recording to areas where the user was given permission.

The snapshots allow users to easily follow the lecture by, for example, identifying the current slide in the slide deck or looking up unfamiliar spellings of words in the lecturer's speech. Furthermore, MaRginalia uses the snapshots to create a history of prior lecture content that allows users to navigate and catch up. Participants considered the prior transcript more useful than the prior slides as it provides more detail. Still, some participants found reading the prior transcript while listening to the speaker challenging. To make this feature more useful, future work can explore providing a visual history that combines elements of the slides as visual cues while summarizing keywords in the transcript for easier catch-up. \rnr{Additionally, we speculate that if the live transcript contains mistakes, it might confuse the user and be distracting. Future work should explore leveraging context in the lecture to reduce transcription errors and highlight keywords to facilitate visual navigation.}

\rnr{In this work, we focused on exploring the in-lecture note-taking experience with MR. Future work should conduct longitudinal and deployment studies with participants in natural lecture settings to quantify the learning benefits of using an MR note-taking system like MaRginalia. Future work can be extended to help users review their notes more easily. For example, such a system might help users better recall the lecture by contextualizing users' notes with relevant data with a timeline (e.g., \textit{ChronoViz} \cite{fouseChronoVizSystemSupporting2011}). Additionally, it could serve as a personalized tutor by assessing users’ understanding through their gaze attention data and the notes they produce.}

\rnr{We focused on university students for this project due to their extensive experience with note-taking in lecture settings and their experience with diverse digital note-taking tools from the flexibility they are offered in adopting digital note-taking tools.}  \rnr{MaRginalia can be extended to scenarios beyond lecture note-taking where slides are not provided beforehand, such as conference and presentation talks.  However, MaRginalia is limited to non-conversational and single-speaker scenarios. To extend to multi-speaker and conversational scenarios, such as board meetings and studio discussions, future work should incorporate meeting transcript techniques (e.g.,~\cite{sonItOkayBe2023}) to help distinguish the speakers and adaptively position the spatial panels to align with the users' attention.} We believe our work opens up an exciting design space for the general intersection of mixed reality and note-taking.

\section{Conclusion}
We introduced MaRginalia, an MR- and tablet-based note-taking system for live in-person lectures that enables note-taking in both MR and directly on the tablet, capturing lecture slides and speaker’s transcripts and navigating the history of the lecture content to catch up on missed content. Through an MR user study, we show that MaRginalia can effectively support note-taking during live lectures. The ability to directly annotate content such as slides enables users to stay engaged with the lecture. Furthermore, easy content transfer proves useful to prevent disorganization of note content lowering user workload, and navigating the history of lectures ensures that users can freely digest information while at the same time keeping up with the live lecture. MaRginalia demonstrates the opportunity for leveraging MR for note-taking and opens up further research for leveraging MR for general productivity and learning.

\begin{acks}
We thank our participants and anonymous reviewers for their invaluable feedback. This research was supported by NSERC Grants RGPIN-2018-04992 and IRCPJ 545100 - 18. 
\end{acks}

\bibliographystyle{ACM-Reference-Format}
\bibliography{Gaze-Note}

\end{document}